\documentclass[aps,twocolumn,superscriptaddress,floatfix]{revtex4-1}

\usepackage{graphicx}
\usepackage{amssymb}
\usepackage{epsfig}
\usepackage[dvipsnames]{xcolor}
\begin{document}

\title{Finite-size effects in the diffusion dynamics
of a glassforming binary mixture with large size ratio}
\author{Vinay Vaibhav}
\affiliation{The Institute of Mathematical Sciences, IV Cross Road, CIT 
Campus, Taramani, Chennai 600 113, Tamil Nadu, India}
\author{J\"urgen Horbach}
\affiliation{Institut f\"ur Theoretische Physik II: Weiche Materie,
Heinrich-Heine-Universit\"at D\"usseldorf, Universit\"atsstra\ss e 1,
40225 D\"usseldorf, Germany}
\author{Pinaki Chaudhuri}
\affiliation{The Institute of Mathematical Sciences, IV Cross Road, CIT 
Campus, Taramani, Chennai 600 113, Tamil Nadu, India}

%%%%%%%%%%%%%%%%%%%%%%%%%%%%%%%%%%%%%%%%%%%%%%%%%%%%%%%%%%%%%%%%%%%%%%%%%%%%
\begin{abstract}
Extensive molecular dynamics (MD) computer simulations of an equimolar
glassforming AB mixture with large size ratio are presented. While
the large A particles show a glass transition around the critical
density of mode coupling theory $\rho_c$, the small B particles
remain mobile with a relatively weak decrease of their self-diffusion
coefficient $D_{\rm B}$ with increasing density.  Surprisingly,
around $\rho_c$, the self-diffusion coefficient of the A particles,
$D_{\rm A}$, also starts to follow a rather weak dependence on
density. We show that this is due to finite-size effects that can
be understood from the analysis of the collective interdiffusion
dynamics.
\end{abstract}                   
%%%%%%%%%%%%%%%%%%%%%%%%%%%%%%%%%%%%%%%%%%%%%%%%%%%%%%%%%%%%%%%%%%%%%%%%%%%

\maketitle

\section{Introduction}
\label{sec1}
Many soft matter systems as well as many biological systems consist
of particles of very different sizes \cite{bechinger2013, weiss2014,
hoefling2013}. These systems may show a glassy dynamics with a
time-scale separation of relaxation processes among the different
constituents. Examples of such systems are glassforming mixtures
of small and large particles that have been studied experimentally
via various colloidal and organic systems \cite{imhof1995_1,
imhof1995_2, kurita2010, blochowicz2012, bierwirth2018, sentjabrskaja2016,
laurati2019} and numerically via hard or soft sphere systems in
computer simulations \cite{moreno2006, horbach2009, xu2012, xu2015,
lazaro2019} as well as in the framework of mode-coupling theory
(MCT) \cite{bosse1987, bosse1995, voigtmann2011}.  A common feature in these
studies is a freezing of the large particles into a glass state
while the small particles remain mobile.  Here, the dynamics of the
small particles is typically associated with anomalous diffusion
on long transient time scales, as reflected, e.g., by a sublinear
growth of the mean-squared displacement $\delta r^2(t)$ as a function
of time $t$, i.e.~$\delta r^2(t) \propto t^\alpha$ with $\alpha<1$.
In computer simulations as well as experiments of disparate-sized
mixtures \cite{kurita2010, blochowicz2012, horbach2009, schnyder2018,
kurzidim2011}, one finds values for the exponent $\alpha$ that in
general depend on the temperature, the total density of the system,
the concentration of small mobile particles, and the interactions
between the particle, especially those between the large and the
small particles. Thus, the values of $\alpha$ are non-universal and
there is typically the lack of a sharp critical point at which one
observes an asymptotic subdiffusive behavior in the longtime limit
with a universal value of the exponent $\alpha$.  This non-universal
behavior can be due to the thermal motion of the particles or soft
interactions between small and large particles.

In a binary mixture of small and large particles, there are the two
corresponding selfdiffusion coefficients $D_{\rm s}$ and $D_{\rm
l}$, respectively, that characterize on one hand the glassy
dynamics of the large particles ($D_{\rm l}$) and on the other hand
the transport of the mobile small particles ($D_{\rm s}$). However,
in addition to these single-particle transport coefficients, there
is also a collective diffusion coefficient, namely the interdiffusion
coefficient $D_{\rm AB}$, that characterizes the mass transport
in the binary mixture \cite{fitts1962, akcasu1997, horbach2007}.
In good approximation, $D_{\rm AB}$ can be often expressed as a
simple linear combination of the selfdiffusion coefficients,
\begin{equation}
D_{\rm AB} = \Phi
\left( x_{\rm l} D_{\rm s} + x_{\rm s} D_{\rm l} \right), 
\label{eq_darken}
\end{equation}
with $x_{\rm l}$ and $x_{\rm s}$ the concentration of the large and
the small particles, respectively, and $\Phi$ the thermodynamic
factor (see below). Equation (\ref{eq_darken}) is often called 
the Darken equation \cite{darken1949} or the Hartley-Crank
equation \cite{hartley1949}.  Computer simulations of glassforming
metallic systems Al-Ni and Zr-Ni with different compositions
\cite{horbach2007, kuhn2014} have shown that Eq.~(\ref{eq_darken})
qualitatively reproduces the temperature dependence of the
interdiffusion coefficient, especially at low temperatures.  

The question of whether the interdiffusion coefficient can be
expressed in terms of the selfdiffusion coefficients has been
extensively discussed in the literature, especially in the context
of (binary) polymer mixtures \cite{akcasu1997, bearman1960,
brochard1986, sillescu1987, akcasu1991, hess1990}.  In this context,
Eq.~(\ref{eq_darken}) is often referred to as the result of a ``fast
mode theory'' \cite{akcasu1997}, because according to Eq.~(\ref{eq_darken})
for a disparate-sized binary mixture the interdiffusion coefficient
would be essentially given by the selfdiffusion coefficient, $D_{\rm
s}$, of the fast mobile species.  In a ``slow mode theory'', however,
the opposite behavior is predicted.  Here, the relation between the
interdiffusion and the selfdiffusion coefficients is given by
\cite{hess1990}
\begin{equation}
D_{\rm AB} = 
\frac{\Phi}{\frac{x_{\rm l}}{D_{\rm s}} + \frac{x_{\rm s}}{D_{\rm l}}}, 
\label{eq_slowm}
\end{equation}
This result can be obtained in the framework of a random phase
approximation \cite{akcasu1997}.  It implies that in a disparate-sized
mixture $D_{\rm AB}$ is dominated by the selfdiffusion coefficient
of the slow species, $D_{\rm l}$.  Note that in the framework of
MCT one also finds that $D_{\rm AB}$ tends to ``follow'' the slow
species such that it always vanishes in a glass state \cite{latz1990}.

For our study, we consider an equimolar binary AB mixture of soft
spheres for which the size ratio of the two species is $\approx
2.85$ and, in addition, the strength of the interaction between AB
pairs is weaker than that between AA and BB pairs. In an earlier
molecular dynamics (MD) simulation study of this system \cite{horbach2009},
it has been demonstrated that on the typical time scale accessible
in the MD simulation the A species falls out-of-equilibrium around
the MCT critical number density which is at $\rho_c = 2.23$,
corresponding to a number density of A particles $\rho_{c}^{\rm
A}=1.115$.  While the A species is in a frozen-in state above
$\rho_c$, the B species remains mobile and there is a relatively
weak decrease of the corresponding selfdiffusion coefficient $D_{\rm
B}$ with increasing density above $\rho_c$. We demonstrate that
Eq.~(\ref{eq_darken}) very well describes the 
density dependence of the interdiffusion coefficients and thus at
high density $D_{\rm AB}$ is proportional to $D_{\rm B}$.  

We show that the approximative proportionality $D_{\rm AB} \propto
D_{\rm B}$ is associated with strong finite-size effects of the
selfdiffusion coefficient of the slow large species, $D_{\rm A}$.
These finite-size effects are due to the relation of $D_{\rm AB}$
to the diffusion coefficient of the centre of mass of species
$\alpha$ (with $\alpha = {\rm A, B}$), $D_{\rm cm}^{(\alpha)}$.
Note that $D_{\rm cm}^{\rm (A)} = D_{\rm cm}^{\rm (B)}$ holds because
the total system's centre of mass is fixed.  As we shall see below,
$D_{\rm cm}^{\rm (A)} \propto D_{\rm AB}/N$, with $N$ the total
number of particles in the system. Thus, the self-diffusion coefficient
of the A species, $D_{\rm A}$, has a finite-size correction $\propto
D_{\rm AB}/N \propto D_{\rm B}/N$ that may be the dominant
contribution to $D_{\rm A}$ for $\rho^{\rm A} \gtrsim \rho_{c}^{\rm
A}$ and small system sizes. Only if one corrects the selfdiffusion
coefficient $D_{\rm A}$ by computing it relative to the center of mass
of the A species, one can extract the true value of $D_{\rm A}$
without the $1/N$ correction. As we argue below, similar features
could be observed in any glassforming system with strong dynamic
heterogeneities.  In such systems, there can be clusters of slow
particles with a relatively fast center-of-mass motion on time
scales where no particle rearrangements inside the cluster occur.
Therefore, our study reveals a common feature in the dynamics of
glassforming liquids.

The rest of the paper is organized as follows: In Sec.~\ref{sec2},
we present the model of the AB mixture, the details of the simulation
and quantities used to analyze the structure and dynamics of the
system. The results of the analysis of structure and dynamics
are given in Sec.~\ref{sec3}, followed by a summary and conclusions in 
Sec.~\ref{sec4}.   

\section{Model and methods}
\label{sec2}
\subsection{Interaction potential and details of the simulation}
\begin{figure}[]
\includegraphics[width=7cm]{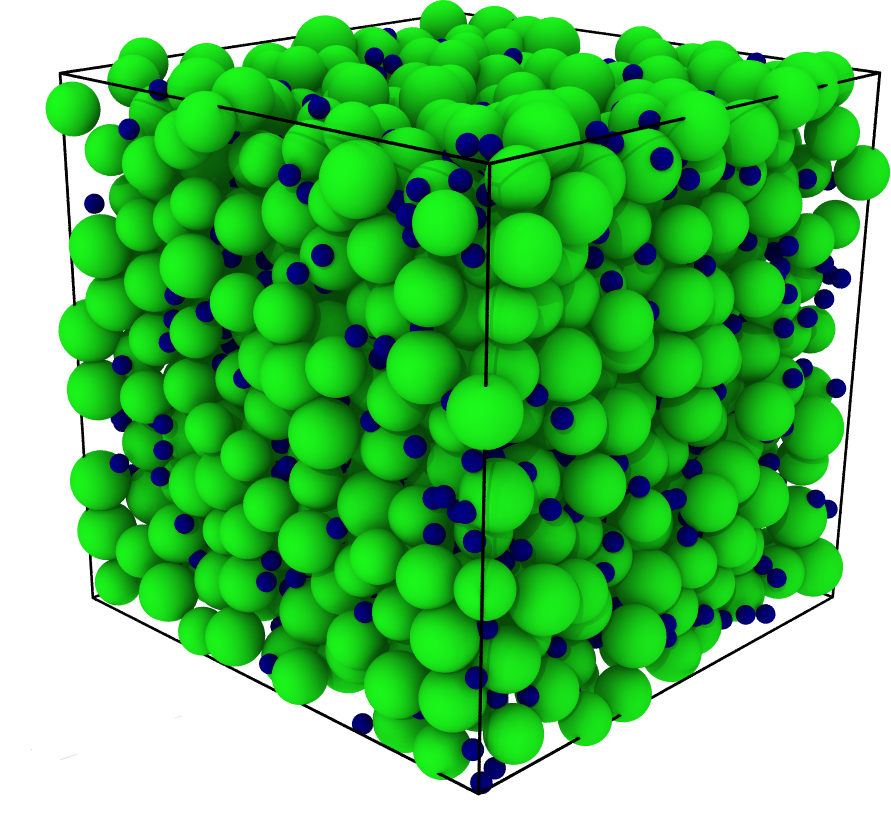}
\caption{Snapshot of the system at the density $\rho = 2.5$. A and
B particles are represented by green and black spheres, respectively.
\label{fig1}}
\end{figure}

The system that we study \cite{horbach2009} is a binary $50-50$
mixture of repulsive particles, where the diameter of the bigger
particles (species A) is sampled from a uniform distribution,
i.e.~$d_{\rm A} \in [0.85,1.15]$, while the diameter of the smaller
particles (species B) is $d_{\rm B} = 0.35$ (see Fig.~\ref{fig1}). The
average size ratio of A and B particles is $\langle d_{\rm A}\rangle/d_{\rm B}
\approx 2.85$ where $\langle d_{\rm A}\rangle \approx 1$. A pair of
particles $\{\alpha,\beta\}$ (with $\alpha = {\rm A, B}$ and
$\beta = {\rm A, B}$), separated by a distance $r$, interacts
via a Weeks-Chandler-Andersen (WCA) potential \cite{weeks1971},
i.e.~a Lennard-Jones potential that is cut off at its minimum and
shifted to zero. To further smoothen the WCA potential, we also add 
a term that provides the continuity of its derivative. Thus, the 
potential is defined by 
\begin{eqnarray}
\label{LJ1}
\textrm{V}_{\alpha\beta}(r) &=& 
u_{\alpha\beta}(r)-u_{\alpha\beta}(R_{c})-\left(r-R_{c}\right)\left. 
\frac{du_{\alpha\beta}}{dr}\right|_{r=R_{c}},\nonumber\\
u_{\alpha\beta}(r) &=& 
4\epsilon_{\alpha\beta}\left[\left(\sigma_{\alpha\beta}/r\right)^{12}-
\left(\sigma_{\alpha\beta}/r\right)^{6}\right]\: ,
\end{eqnarray}
for $r<R_{c} = 2^{1/6} \sigma_{\alpha\beta}$, else
$\textrm{V}_{\alpha\beta}(r) = 0$, with $\alpha, \beta$ as particle
index. Here, $\sigma_{\alpha\beta} = (d_{\alpha}+d_{\beta})/2$ and
$\epsilon_{\alpha\beta} = \epsilon = 1.0$ if both interacting
particles are of the same type, else $\sigma_{\alpha\beta} =
(1.00+0.35)/2$ and $\epsilon_{\alpha\beta} = 0.1$.  All particles
have the same mass $m_{\rm A} = m_{\rm B} = m = 1$.  In the following,
length, energy, and time are measured in units of $\langle d_{\rm
A} \rangle$, $\epsilon$, and $\tau_{\rm WCA} = [m \langle d_{\rm
A} \rangle ^2/\epsilon]^{1/2}$, respectively.

%%%%%%%%%
%%%%%%%
Using LAMMPS \cite{plimpton1995}, extensive molecular dynamics (MD)
simulations are performed for systems with $N = 1000$, 2000, and
4000 particles, placed in a three-dimensional cubic box with periodic
boundary conditions.  The equations of motion are integrated using
the velocity form of the Verlet algorithm \cite{allen2017} with a
time step $dt = 0.00075 \, \tau_{\rm WCA}$.  The simulations are
done for different number densities in the range $2.1 \le \rho \le
3.5$ at the temperature $T = 2/3$. For each density, 30 independent
samples are simulated.  During the equilibration of the samples,
the temperature is kept constant using a dissipative particle
dynamics (DPD) thermostat \cite{soddemann2003} where the damping
coefficient is set to 1.0.  For the high densities, $\rho \ge 2.25$,
the MD simulations are combined with the swap Monte Carlo (SMC)
algorithm \cite{grigera2001, berthier2019} to obtain well-annealed
samples.  In a trial SMC move, one randomly selects a pair of
particles, exchanges their diameters, and accepts or rejects this
move according to a Metropolis criterion. In our scheme, only the
diameters of the A particles are swapped. Every 100 MD steps, $N_{\rm
A}$ trial SMC moves are done.  The longest equilibration runs
with the hybrid MD-SMC method were over $6 \times 10^8$ time steps.
After the equilibration, the thermostat as well as the SMC are switched 
off, performing the production runs in the microcanonical ensemble.

A snapshot of the system at the density $\rho = 2.5$ is shown in
Fig.~\ref{fig1}. From this snapshot, one can infer that the large
A particles form a close-packed structure while the small B particles
can explore the free volume between the A particles.

\subsection{Structural and dynamic properties}
In this section, we define the correlation functions and transport
coefficients that we use to analyse the simulation results for our
AB mixture. A central static correlation function for our analysis
is the concentration-concentration structure factor $S_{cc}(q)$.
In the limit $q\to 0$, this function is related to the thermodynamic
factor $\Phi$ in Eq.~(\ref{eq_darken}). After having introduced
$S_{cc}(q)$ and its relation to $\Phi$, we show how the selfdiffusion
as well as the interdiffusion coefficients can be computed via
Einstein relations, i.e.~via long-time limits of mean-squared
displacements.

We consider an AB mixture that contains a total number of $N =
N_{\rm A} + N_{\rm B}$ particles. Thus, the concentration of A
and B particles is given by $x_{\rm A} = N_{\rm A}/N$ and
$x_{\rm B} = N_{\rm B}/N$, respectively. The local number density in
reciprocal space for particles of type $\alpha$ can be defined as
follows \cite{hansen1986}:
\begin{equation}
\rho_{\alpha} (\vec{q}) = 
\sum_{j=1}^{N_{\alpha}} \exp \left( i \vec{q} \cdot \vec{r}_j \right)
\label{eq_rhoq}
\end{equation}
with $\vec{q}$ the wavevector and $\vec{r}_j$ the position of the
$j$'th particle of type $\alpha$. The autocorrelation functions of
the density variables, as defined by Eq.~(\ref{eq_rhoq}), are the
partial structure factors \cite{hansen1986},
\begin{eqnarray}
S_{\alpha \beta}(q) & = & 
\frac{1}{N} \left\langle 
\rho_{\alpha}(\vec{q}) \rho_{\beta}(- \vec{q}) \right\rangle \nonumber \\  
 & = & \frac{1}{N}  \sum_{j=1}^{N_\alpha}  \sum_{k=1}^{N_\beta} 
\left\langle 
\exp\left[ - i \vec{q} \cdot \left( \vec{r}_j - \vec{r}_k \right) \right] \right\rangle,
\label{eq_spart}
\end{eqnarray}
where $\langle . \rangle$ indicates an ensemble average. Note that
we assume in Eq.~(\ref{eq_spart}) that the system is isotropic and
thus the partial structure factors only depend on the magnitude of the
wavevector, $q$.

From the densities (\ref{eq_rhoq}), we can introduce local concentration
fluctuation variables \cite{hansen1986} as
\begin{equation}
c_\alpha (\vec{q}) = 
\rho_\alpha(\vec{q}) - 
x_\alpha \left( \rho_{\rm A}(\vec{q}) 
+ \rho_{\rm B}(\vec{q}) \right) \, ,
\end{equation}
describing the local deviation from a homogeneous distribution of
particles of type $\alpha$. Since $c_{\rm A} + c_{\rm B} = 0$, the
concentration variables are not independent of each other and it
suffices to define {\it one} concentration-concentration structure
factor for the binary mixture as
\begin{equation}
S_{cc}(q) = 
\frac{1}{N} \left\langle c_{\rm A}(\vec{q}) 
c_{\rm A}(-\vec{q}) \right\rangle \, .
\end{equation}
This function can be also expressed as a linear combination of the 
partial structure factors,
\begin{equation}
S_{cc}(q) =
x_B^2S_{\rm AA}(q)+x_{\rm A}^2 S_{\rm BB}(q) -
2x_{\rm A} x_{\rm B} S_{\rm AB}(q).   
\end{equation}
In the limit $q\to \infty$, $S_{cc}(q)$ approaches $x_{\rm A} x_{\rm
B}$, corresponding to the concentration-concentration structure
factor of an ideal binary mixture. In the limit $q\to 0$, $S_{cc}(q)$
is related to the concentration susceptibility and thus to the
second derivative of the Gibbs free energy $G$ with respect to
$x_{\rm A}$ and $x_{\rm B}$ via
\begin{equation}
\Phi = \frac{x_{\rm A} x_{\rm B}}{k_{\rm B} T} 
\frac{\partial^2 G}{\partial x_{\rm A} x_{\rm B}}
= \frac{x_{\rm A} x_{\rm B}}{S(q=0)} \, ,
\label{eq_phi}
\end{equation}
with $k_{\rm B}$ the Boltzmann constant. Below, we use Eq.~(\ref{eq_phi}) 
to compute the thermodynamic factor from an extrapolation of $S_{cc}(q)$
to $q=0$. 

Now we introduce quantities that characterize dynamic properties 
of the AB mixture. At a single-particle level, we consider the
incoherent intermediate scattering function $F_{\rm s}^{\alpha}(q,t)$
of a tagged particle of type $\alpha$. This is the correlation function   
of the time-displaced one-particle density and defined by \cite{hansen1986}
\begin{equation}
\label{eq_fsqt}
F_{\rm s}^{\alpha}(q,t) = 
\frac{1}{N_{\alpha}}  \sum_{j=1}^{N_\alpha} 
\left\langle \exp \left[ -i \vec{q}\cdot 
\left( \vec{r}_j(t) - \vec{r}_j(0) \right) \right] 
\right\rangle.
\end{equation}
Of special interest for the analysis of the dynamics of glassforming
liquids is the decay of $F_{\rm s}^{\alpha}(q,t)$ as a function of
time around values of $q$ corresponding to the location of the first
peak of the static structure factor, $q_{\rm max}$. This is due to
the fact that slowing down of the structural relaxation of the
glassforming liquid is associated with the cage effect \cite{binder2011}
and the typical size of a cage is of the order of $2 \pi/q_{\rm
max}$. In our case, the A particles exhibit a typical glassy dynamics
and the first peak of $S_{\rm AA}(q)$ is at $q_{\rm max} \approx
6$.  Thus, below we consider $F_{\rm s}^{\alpha}(q,t)$ at this value
of $q$.

We determine the selfdiffusion coefficient of a tagged particle of
type $\alpha$ from the mean-squared displacement (MSD), defined by
\begin{equation}
\label{eq_msd}
\left\langle \delta r^2_{\alpha}(t) \right\rangle =  
\frac{1}{N_{\alpha}}  \sum_{j=1}^{N_\alpha} 
\left\langle \left( \vec{r}_j(t) - \vec{r}_j(0) \right)^2 \right\rangle,
\end{equation}
The corresponding selfdiffusion coefficient $D_\alpha$ is obtained from the
long-time limit of $\left\langle \delta r^2_{\alpha}(t) \right\rangle$
via the Einstein relation \cite{hansen1986, binder2011}
\begin{equation}
D_{\alpha} = 
\lim_{t \to \infty} 
\frac{\left\langle \delta \tilde{r}^2_{\alpha}(t) \right\rangle}{6t}.
\label{eq_ds}
\end{equation}

We also compute modified versions of the incoherent intermediate
scattering function and the MSD where we replace the coordinates
of the particles $\vec{r}_j(t)$ in Eqs.~(\ref{eq_fsqt}) and
(\ref{eq_msd}) by
\begin{equation}
\vec{r}^{\; \prime}_j(t) = \vec{r}_j(t) - \vec{R}_{\rm A}(t) \, . 
\label{eq_rprime}
\end{equation}   
In this equation, $\vec{R}_{\rm A}(t)$ is the center-of-mass coordinate of the
A particles at time $t$,
\begin{equation}
\vec{R}_{\rm A}(t) =
\frac{1}{N_{\rm A}} \sum_{j=1}^{N_{\rm A}} \vec{r}_j(t) \, ,
\end{equation}
with $\vec{r}_j$ the position of the $j$'th particle of type $A$.
The purpose of calculating $F_{\rm s}^\alpha(q,t)$ 
and $\left\langle \delta r^2_{\alpha}(t) \right\rangle$ with the 
center-of-mass-corrected coordinates (\ref{eq_rprime}) 
will become clear below.

The interdiffusion coefficient $D_{\rm AB}$ can be also calculated
from an Einstein relation, i.e.~from the long-time limit of a 
MSD. In this case, the MSD
of the variable $\vec{R}_{\rm A}(t)$ has to be considered,
\begin{equation}
\left\langle \delta R^2(t) \right\rangle_{\rm cm} =
\left\langle \left[ 
\vec{R}_{\rm A}(t) - \vec{R}_{\rm A}(0) \right]^2
\right\rangle \, .
\label{eq_msdcm}
\end{equation}
Then, the interdiffusion coefficient is given by \cite{horbach2007}
\begin{equation}
D_{\rm AB} = \Phi L \, ,
\label{eq_dab}
\end{equation}
where $\Phi$ is the thermodynamic factor, as defined by Eq.~(\ref{eq_phi}),
and the Onsager coefficient
\begin{equation}
L = 
\left[ 1+ \frac{x_{\rm A}}{x_{\rm B}} \right]^2
N x_{\rm A} x_{\rm B}
\lim_{t\to \infty}
\frac{\left\langle \delta R^2(t) \right\rangle_{\rm cm}}{6t} \, ,
\end{equation}
describes the kinetic part of $D_{\rm AB}$. 

One can also define a center-of-mass diffusion coefficient for each
species which, for species A, is given by
\begin{equation}
D_{\rm A}^{\rm cm} =
\lim_{t \to \infty}
\frac{\left\langle \delta{R^2(t)} \right\rangle_{\rm cm}}{6t} \, .
\label{eq_dcm}
\end{equation}
For the mixture that is being studied, where $x_{\rm A} = x_{\rm
B} = 1/2$, we therefore obtain $D_{\rm AB} = \Phi N D_{\rm A}^{\rm
cm}$ and $L = N D_{\rm A}^{\rm cm}$.

Formally, the interdiffusion coefficient $D_{\rm AB}$ can be written
as a linear combination of the selfdiffusion coefficients,
\begin{equation}
D_{\rm AB} = \Phi S 
\left( x_{\rm A} D_{\rm B} + x_{\rm B} D_{\rm A} \right) \, ,
\label{eq_dabdarken}
\end{equation}
where the ``Manning factor'' \cite{manning1961} $S$ contains all
the cross correlations \cite{horbach2007} that contribute to $L$. 
The value $S = 1$ implies vanishing cross correlations
\cite{akcasu1997, horbach2007}. In this case, the Darken equation
(\ref{eq_darken}) holds.

\section{Results}
\label{sec3}
\subsection{Structural and thermodynamic properties}
\begin{figure}[]
\includegraphics[width=8cm]{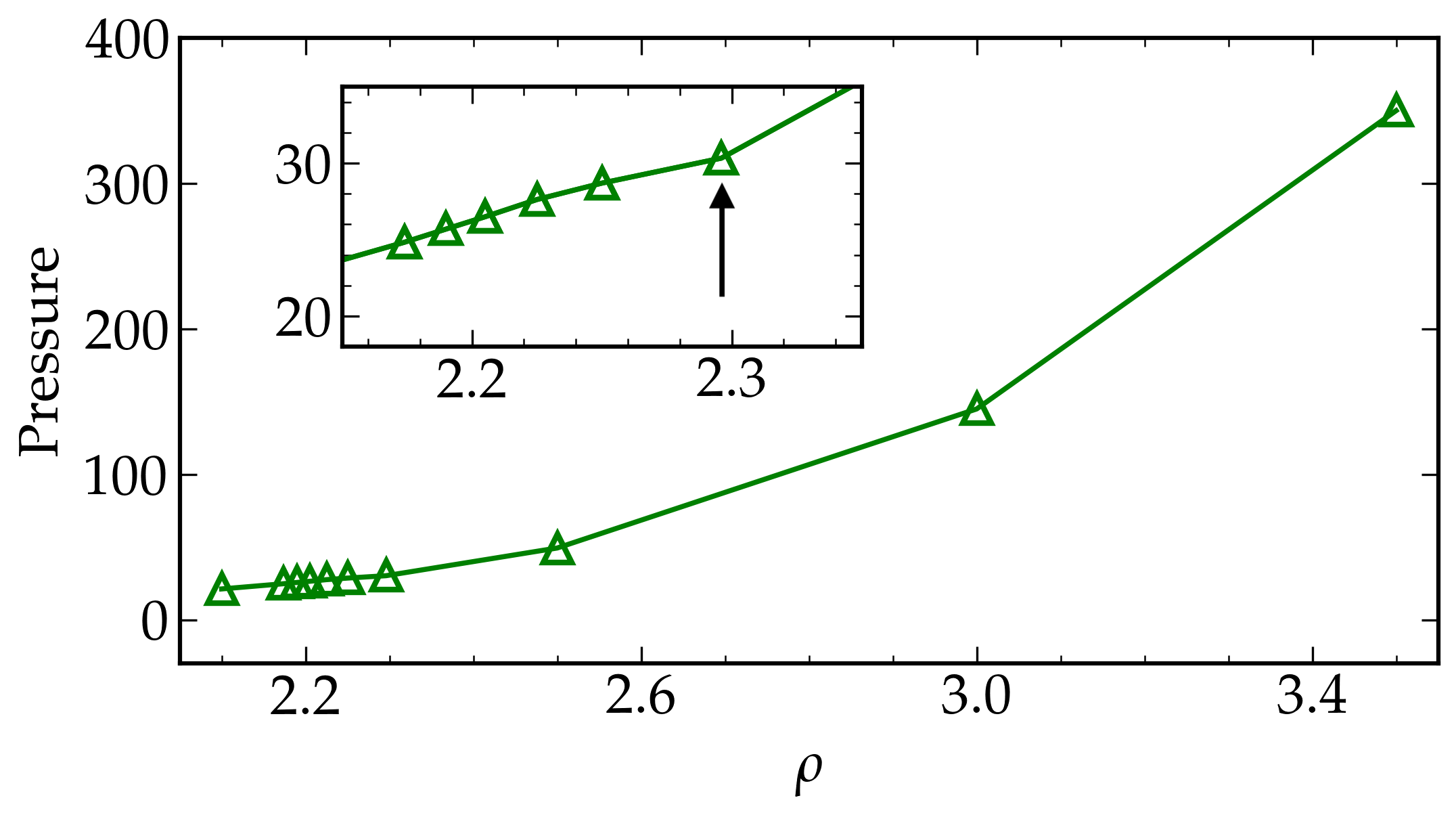}
\caption{Pressure $P$ as a function of density. The inset zooms in 
the region around $\rho = 2.296$ (this density is indicated by an arrow). 
\label{fig2}}
\end{figure}
With the aid of the SMC technique, we are able to obtain well-annealed
samples at very high densities that are above the critical MCT
density $\rho_c \approx 2.23$ \cite{horbach2009}. However, at such
high densities, one may expect that at least for the A particles,
the thermodynamic equilibrium is an ordered crystalline phase.
Although the large polydispersity of the A particles in our model
suppresses crystallization to some extent, the use of the SMC
technique tends to also accelerate the formation of crystalline
clusters and therefore we check especially the high-density samples
whether they are purely amorphous structures and thus free of any
crystalline clusters. To this end, we measure static structure
factors and analyse the samples in terms of local bond order
parameters.  Furthermore, we determine the thermodynamic factor
$\Phi$ from the concentration-concentration structure factor
$S_{cc}(q)$.

The pressure $P$ increases monotonously with increasing density
$\rho$ (Fig.~\ref{fig2}).  However, the lines connecting the points
between $\rho = 2.25$ and $\rho = 2.296$ in $P(\rho)$ indicate a
slight change of the slope (cf.~the inset of Fig.~\ref{fig2}). This
could be due to a liquid-solid coexistence, occurring around these
densities.  And indeed our analysis for the density $\rho = 2.296$
(see below) indicates the occurrence of crystalline clusters
in some of the samples at that density.

\begin{figure}[]
\includegraphics[width=8cm]{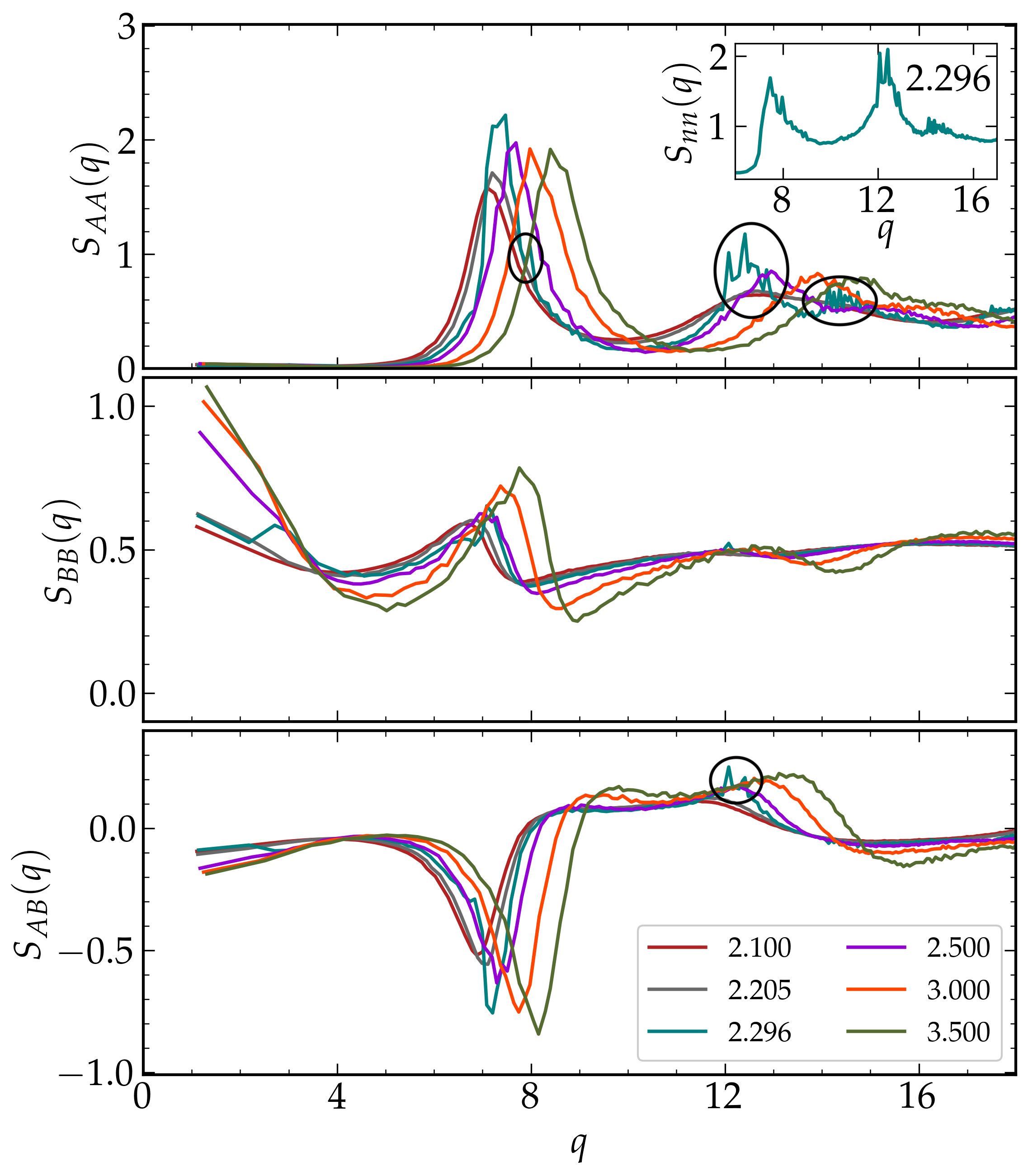}
\includegraphics[width=8cm]{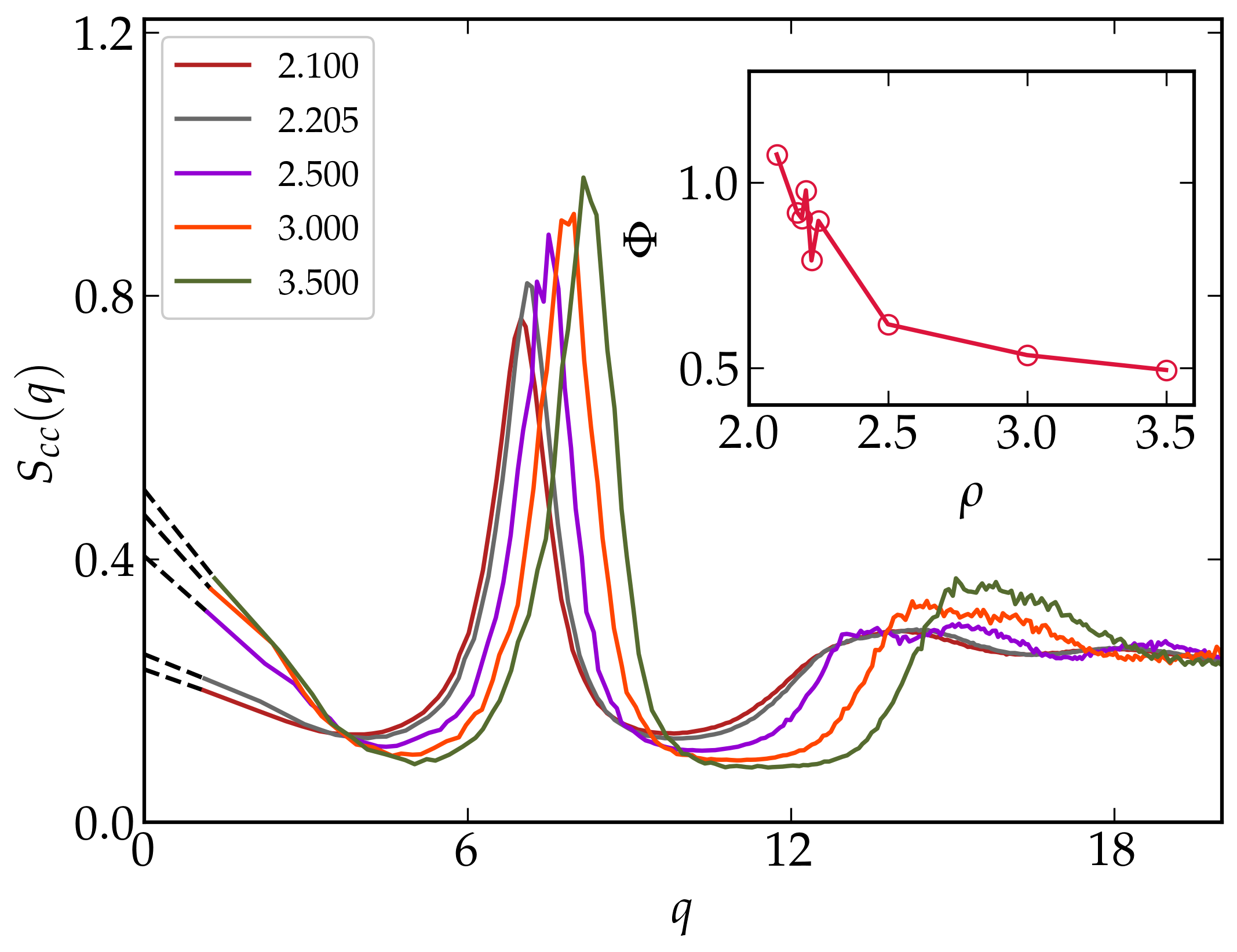}
\caption{{\bf Top panels:} Partial structure factors $S_{\rm AA}(q)$,
$S_{\rm BB}(q)$, and $S_{\rm AB}(q)$ at different densities.  Emerging
Bragg peaks are indicated by circles. The inset of the plot of
$S_{\rm AA}(q)$ shows the total structure factor $S_{nn}(q)$ at the
density $\rho = 2.296$. {\bf Bottom panel:} Concentration-concentration
structure factor $S_{cc}(q)$ at different densities. The dashed
lines indicate the extrapolation to $q=0$ via a fit function (see
text). The thermodynamic factor $\Phi$, as obtained from the fitted
$S(q=0)$, is shown in the inset as a function of density. \label{fig3}}
\end{figure}

To quantify the structural changes in our AB mixture with increasing
density, we now consider the partial structure factors
$S_{\alpha\beta}(q)$. The top panels of Fig.~\ref{fig3} show these
functions for different values of $\rho$.  With increasing $\rho$,
the first peak in both $S_{\rm AA}(q)$ and $S_{\rm BB}(q)$ shifts
to larger $q$ as the inter-particle separation decreases. In between,
at the density $\rho = 2.296$, $S_{\rm AA}(q)$ shows signatures of
possible formation of local crystallites, with discrete spikes being
clearly visible. Although less prominent, this is also reflected
in the cross correlation, $S_{\rm AB}(q)$, as well as in the total
structure factor $S_{nn}(q) = S_{\rm AA}(q) + S_{\rm BB}(q) + 2
S_{\rm AB}(q)$ that is shown in the inset of the plot of $S_{\rm
AA}(q)$. However, for densities higher than $\rho = 2.296$ there
is no sign of any Bragg peaks, suggesting that the samples at these
high densities is purely amorphous. In fact, this is confirmed by
our analysis of the samples in terms of local bond order parameters
(see below).

In the bottom panel of Fig.~\ref{fig3}, we show how $S_{cc}(q)$
varies with increasing density. We extract $S_{\rm cc}(0)$ by
extrapolating $S_{cc}(q)$ to $q = 0$, using the fit function $f(q)
= S_{cc}(0) [1 - A\, q^2 + B\, q^4]$ (with $S_{cc}(0)$,$A$, and $B$
being fit parameters). For the fits (dashed lines), we have only
taken into account the data for $q \le 2.0$. Then, from $S_{cc}(0)$
we compute the thermodynamic factor $\Phi$ via Eq.~(\ref{eq_phi}).
The variation of  $\Phi$ is shown in the inset.  We observe that $\Phi$ decreases from
a value of about 1.0 at $\rho = 2.0$ to a value of about 0.5 at the
highest considered density, $\rho = 3.5$.  This rather weak variation
of $\Phi$ over a broad range of densities also implies that the
thermodynamic factor does not strongly affect the density dependence
of the interdiffusion coefficient, which we discuss further below.

\begin{figure}[]
\includegraphics[width=8cm]{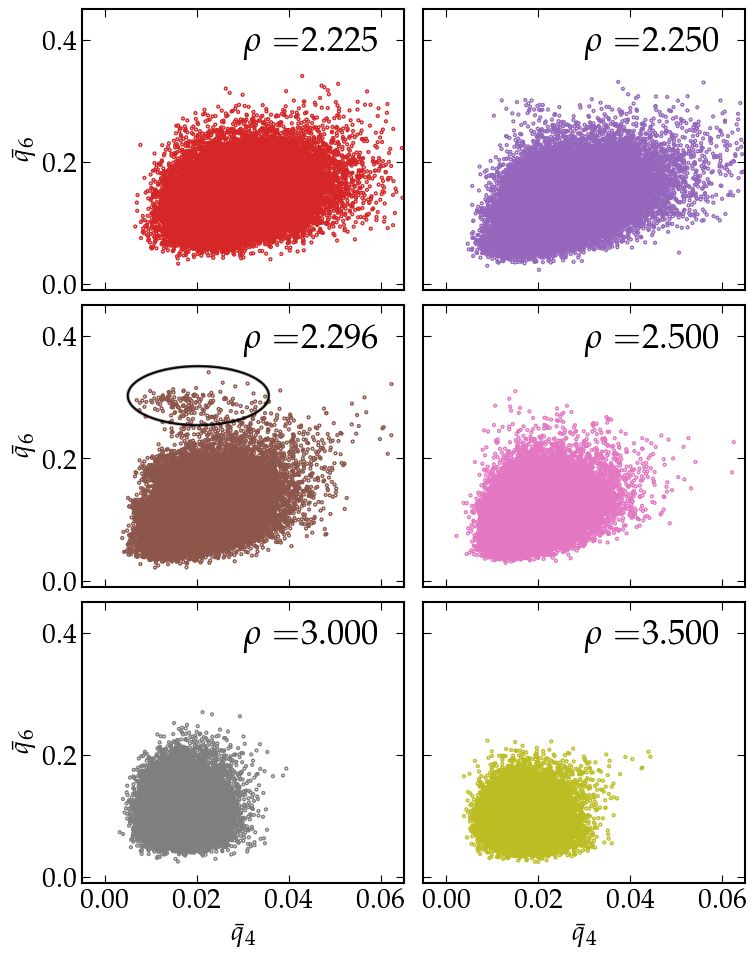}
\caption{$\bar{q}_4-\bar{q}_6$ plot for system at different marked
densities (calculation with cutoff). The closed line in the plot 
for $\rho = 2.296$ indicates the values of $\bar{q}_4-\bar{q}_6$ pairs
that correspond to crystalline BCC clusters (see text). \label{fig4}}
\end{figure}

Our analysis to check for local crystalline order is based on the
local bond order parameters that have been proposed by Steinhardt
{\it et al.}~\cite{steinhardt1983}. For such analysis, at first we
identify the nearest neighbours for each particle within the cutoff
$1.5$ ($N_b(i)$ is the number of such neighbours for each particle
$i$). After this, for each particle, a complex local orientational
order vector $q_{lm}(i)$ with $(2l+1)$ components is constructed
using the following definition:
\begin{equation}
\label{qlm}
q_{lm}(i) =
\frac{1}{N_b(i)} \sum_{j=1}^{N_b(i)} Y_{lm}(\vec{r}_{ij}).
\end{equation}
Here, $Y_{lm}(\vec{r}_{ij})$ is the spherical harmonic of degree
$l$ and order $m$. As per definition, $l$ is always a non-negative
integer and $m$ can take the integral values from $m = -l$ to $m =
l$ for a given value of $l$. Also, $\vec{r}_{ij}$ is the vector
from the particle $i$ to particle $j$. Now, as suggested in
Ref.~\cite{dellago2008}, the locally averaged quantity $\bar{q}_{lm}(i)$
is calculated for each particle $i$, using
\begin{equation}
\bar{q}_{lm}(i) = 
\frac{1}{\tilde{N}_b(i)} \sum_{k=0}^{\tilde{N}_b(i)} q_{lm}(k),
\end{equation}
where averaging (summation from $k=0$ to $\tilde{N}_b(i)$) has been
done over the neighboring particles of $i$ (as above defined, based
on the cutoff) and particle $i$ itself. Such averaging procedure
takes into account even the information of the structure beyond the
cutoff and has shown better identification of crystal structures
in simulations. Using these averaged form of local bond order vector
components a norm $\bar{q}_{l}(i)$ is defined for each particle:
\begin{equation}
\bar{q}_{l}(i) = 
\sqrt{\frac{4\pi}{2l+1} \sum_{m=-l}^{l} |\bar{q}_{lm}(i)|^2}.
\end{equation}
Depending on the definition of the spherical harmonics used, the
factor $\frac{4\pi}{2l+1}$ before summation in the above equation, is
used or not used. This quantity $\bar{q}_{l}(i)$ is sensitive to
different crystal structure systems depending on the choice of $l$.
Specially, bond orientational order parameter for $l = 4$ and $l =
6$ have been used to identify structures similar to cubic and
hexagonal systems. We have calculated the correlation
between $\bar{q}_{4}$ and $\bar{q}_{6}$ for the A particles in our
system.

The measured values of $\bar{q}_{4}$ and $\bar{q}_{6}$, for each
particle across several representative configurations, can be
visualized in the form of a scatter plot in the $\bar{q}_4$-$\bar{q}_6$
plane, following Ref.~\cite{dellago2008}, to check if local
environments exhibit any formation of ordered structures. This is
shown in Fig.~\ref{fig4}, for a few densities across the range that
we have studied, using configurations sampled from $30$ independent
trajectories. First thing to notice is that, across all the densities,
the values of $\bar{q}_4$ are small ($<0.06$) and in fact shrink
with increasing density. Similarly for $\bar{q}_6$, the numbers are
in the liquid-like regime, except for $\rho=2.296$ where some pockets
of BCC-like structures seem to be visible for some clusters within
some trajectories, if we compare with observations reported in
Ref.~\cite{dellago2008}.

To summarize, the structural changes with increasing density are
on expected lines. There is some hint of occurrence of a small
number of local crystallites, in the vicinity of the mode coupling
density. However, when supercooled to higher densities, the system
remains disordered, even when equilibriated via SMC.

\subsection{Diffusion dynamics}
We will now discuss the dynamic properties of the mixture.  As
mentioned earlier, for the model system that we are studying, the
larger A species are known to exhibit a mode coupling transition
around $\rho_c \approx 2.23$ \cite{horbach2009}.  For $\rho \gtrsim
\rho_c$, the A particles are essentially in a frozen-in configuration
on the diffusive time scale of the B particles.  However,  the
diffusivity of the B particles also continues to decrease with increasing
density such that they are expected to be in an arrested state at
very large densities.

\begin{figure}[]
\includegraphics[width=8cm]{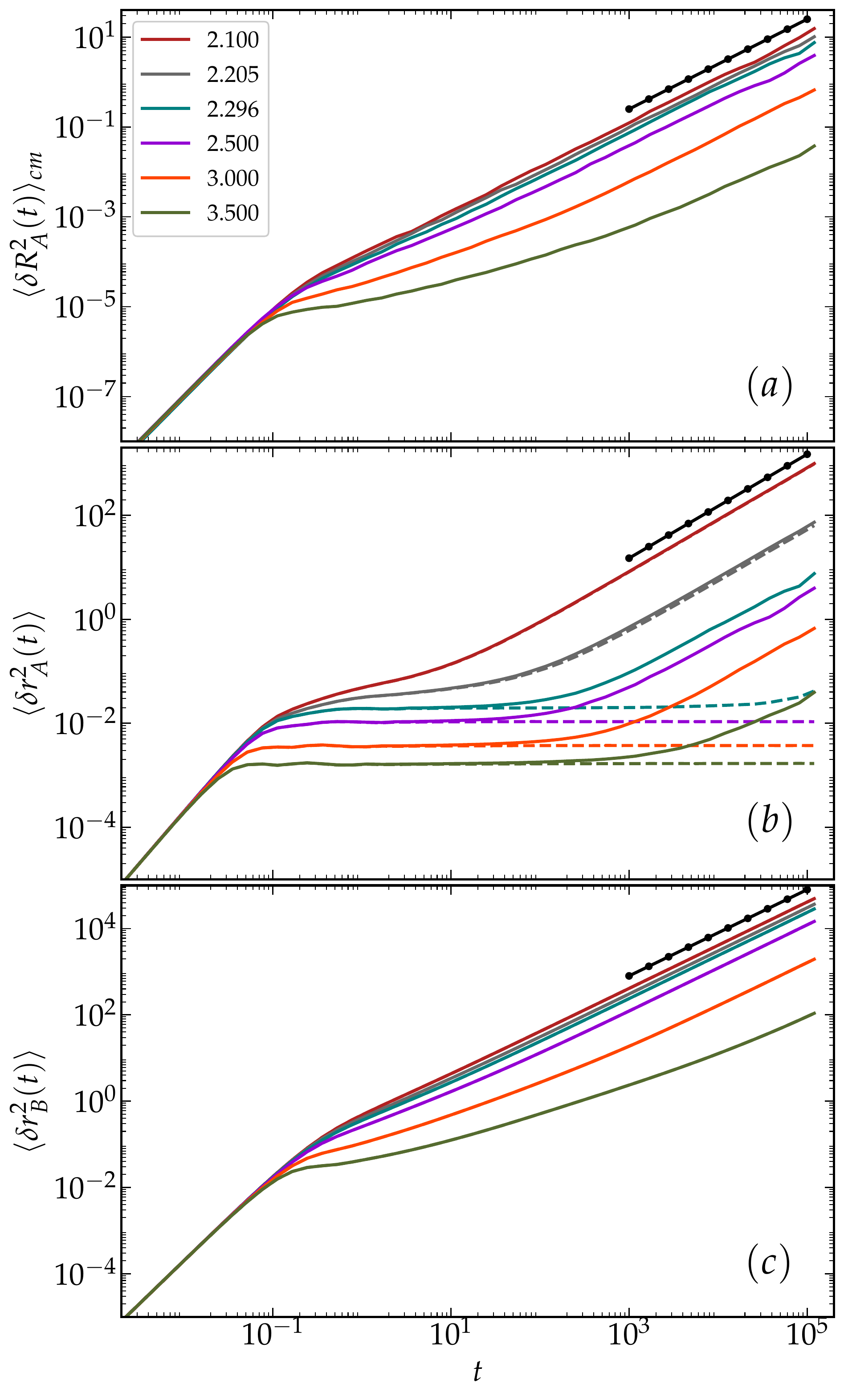}
\caption{(a) Center-of-mass MSD, $\left\langle \delta R^2(t)
\right\rangle_{\rm cm}$, for different densities. (b) Single-particle
MSDs of A species and (c) of B species for the same densities. In
b) and c), solid and dotted lines represent calculations with and
without the center-of-mass correction, respectively (see text).  In
all sub-plots, the solid lines with dots are straight lines $\propto
t$ to indicate the diffusional regime of the MSDs at long times.
\label{fig5}}
\end{figure}
\begin{figure}[]
\includegraphics[width=8cm]{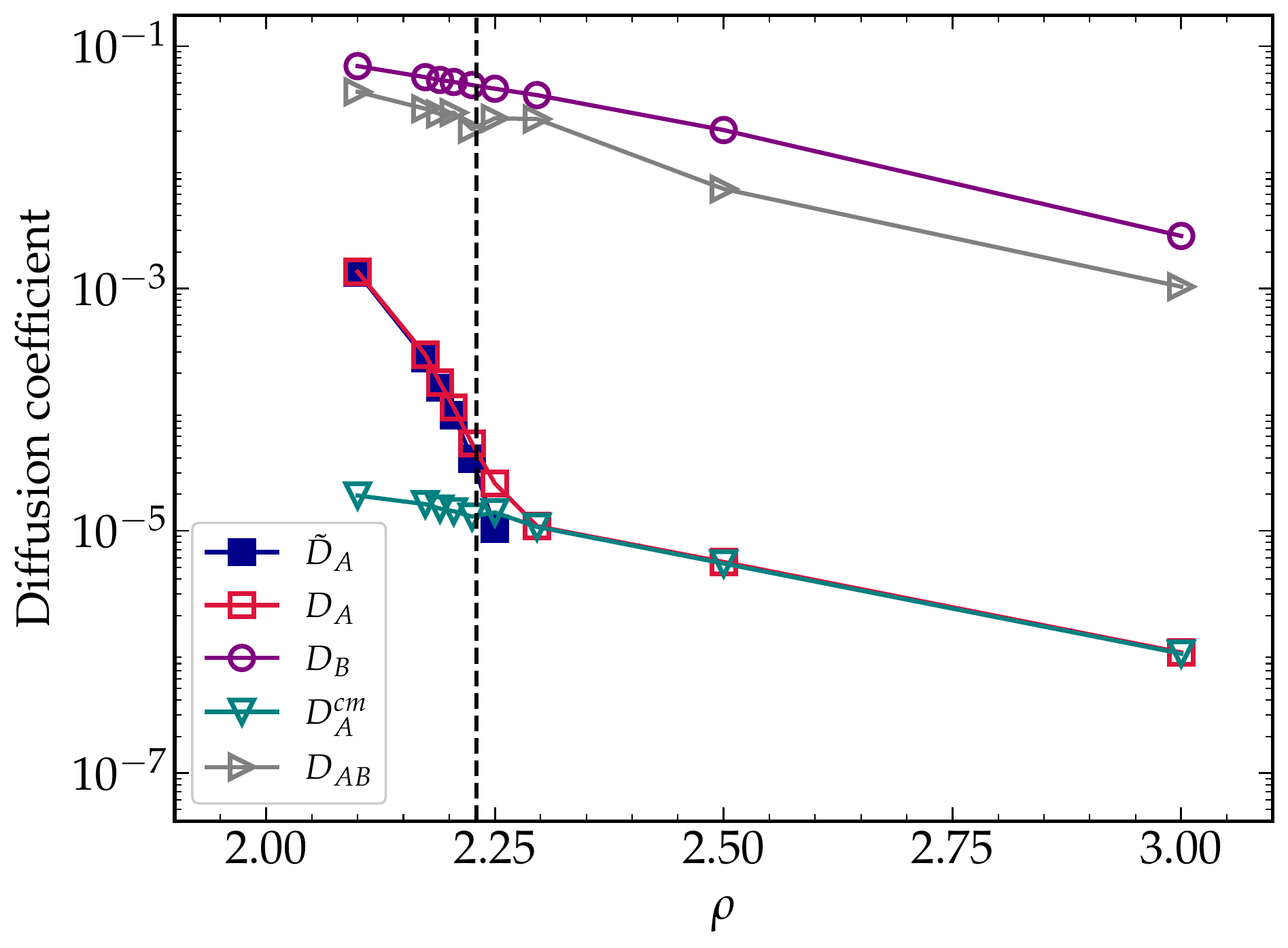}
\includegraphics[width=8cm]{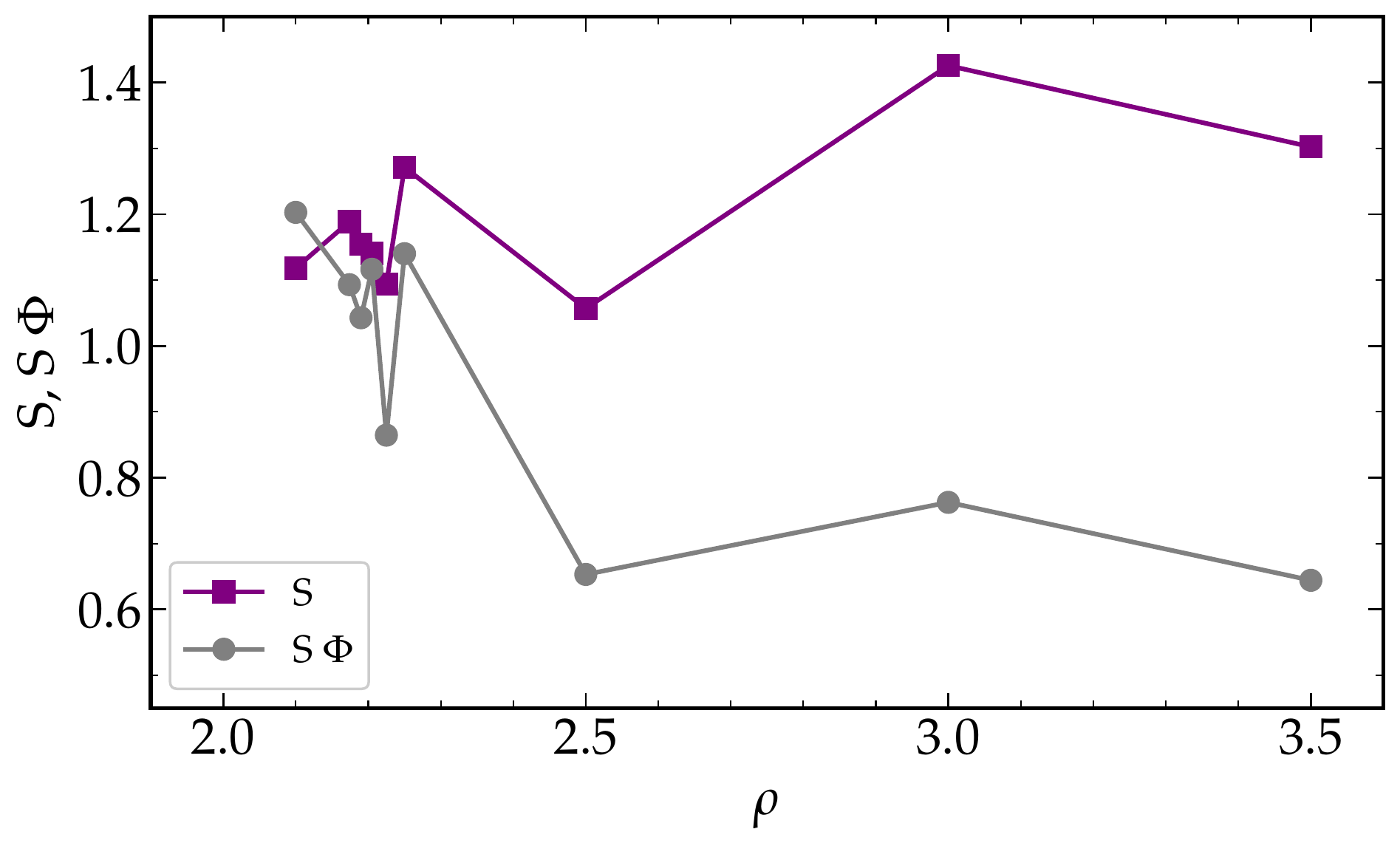}
\caption{{\bf Top panel:} Selfdiffusion coefficients $D_{\rm A}$
and $D_{\rm B}$, the center-of-mass diffusion coefficient $D_{\rm
A}^{\rm cm}$  and the interdiffusion coeffient $D_{\rm AB}$ as a
function of density $\rho$. Also shown is the
rectified self-diffusion coefficient $\tilde{D}_{\rm A}$ (see text).
The vertical dashed line marks the
previously estimated critical MCT density for $D_{\rm A}$ at
$\rho_c=2.23$.  {\bf Bottom panel:} Manning factor $S$ and the
product $S \Phi$ as a function of density. \label{fig6}}
\end{figure}

In the following, our main objective is to study the density
dependence of the interdiffusion coefficient. To this end, one has
to monitor the trajectory of the center of mass of the A species
and the corresponding time evolution of $\langle \delta R^2(t)
\rangle_{\rm cm}$, as defined by Eq.~(\ref{eq_msdcm}), needs to be
computed.  This MSD is shown in Fig.~\ref{fig5}a for different
densities.  For the same densities, Figs.~\ref{fig5}b and \ref{fig5}c
display the single-particle MSDs, $\langle \delta r^2_{\alpha}(t)
\rangle$ (see Eq.~(\ref{eq_msd})), of the A and B particles,
respectively.  For the A species, $\langle \delta r^2_{\rm A}(t)
\rangle$ shows the behaviour that is expected for a typical
glassforming liquid. There is a short-time ballistic
regime ($\propto t^2$) and a long-time diffusive regime ($\propto
t$). In between these two regimes, there is a plateau-like region
that becomes more pronounced and broader with increasing density.
The latter regime is due to the intermediate caging of the particles.
Importantly, we note that even for $\rho > \rho_c$, the MSD of A
particles still displays diffusive dynamics at long times. Below,
we show that this feature is a finite-size effect, i.e.~the observed
diffusive regime for $\rho > \rho_c$ shifts to longer time scales
with increasing system size.  For the case of B species, the shape
of the MSD, i.e.~of $\langle \delta r^2_{\rm B}(t) \rangle$, is
very different from that of the A particles, with no intermediate
plateau at all. We will also discuss this, below.  Now, if we look
at the MSD curves for the center-of-mass motion, it is evident that
they bear close resemblance with that of the B species, the reason
for which will be evident once we analyse the diffusion coefficients.

The long-time dynamical behaviour is well characterized by measuring
the respective diffusion coefficients, measured from the corresponding
MSD data, as defined in Eqs.~(\ref{eq_ds}), (\ref{eq_dab}), and
(\ref{eq_dcm}).  The density variation of the measured single-particle
diffusion coefficients, $D_{\rm A}$ and $D_{\rm B}$, the center-of-mass
diffusion coefficient $D_{\rm A}^{\rm cm}$ as well as the interdiffusion
coeffient $D_{\rm AB}$ are shown in Fig.~\ref{fig6}.  First, we
note that the diffusion coefficient of the smaller B particles
remains finite beyond $\rho_c$ (indicated via the dashed vertical
line), consistent with previous study \cite{horbach2009}.  Now, if
we look at the data for $D_{\rm A}$, we observe that it decreases
as it approaches $\rho_c$, as reported earlier, and beyond that
there seems to be a separate branch of weaker decrease with density.
Interestingly, for $\rho > \rho_c$, the measured center-of-mass
diffusion coefficient,  $D_{\rm A}^{\rm cm}$, shows exactly the
same density dependence.  As already noted in the previous paragraph,
the measured interdiffusion coefficient has a density dependence
which resembles that of the diffusivity of B species, over the
entire density range.

We now analyse the motion of A particles for $\rho > \rho_c$. Since
the density dependence of $D_{\rm A}$, in this density regime,
matches that of $D_{\rm A}^{\rm cm}$, it implies that the observed
motion of the individual A particles is actually due to the motion
of the center of mass of the A species.  To disentangle that, we
compute the MSD of the A particles by shifting to the frame of
reference of the population's center of mass, see Eq.~(\ref{eq_rprime}).
The redefined MSDs are plotted in the top panel of Fig.~\ref{fig5},
using dashed lines. We observe that following this rectification,
there is a significant change in the MSD for $\rho > \rho_c$. The
rectified MSDs, in that regime, exhibit a prolonged plateau over
the time window of our observation implying that in the frame of
reference of the center of mass, the particles are essentially caged
and there is hardly any cage-breaking, especially for $\rho \geq
2.5$. Therefore, we can conclude that only due to the motion of the
center of mass, deviations from the plateau are exhibited in the
bare MSDs at these densities. At small enough density ($\rho=2.1$),
this rectification is not observed, and there is a mild rectification
for $\rho=2.205$.  The rectified selfdiffusion coefficient of the
A particles is also shown in Fig.~\ref{fig6}, from which it is
evident that the rectified $D_{\rm A}$, labelled as $\tilde{D}_{\rm A}$, 
sharply decreases around $\rho_c$. 

\begin{figure}[t]
\includegraphics[width=8cm]{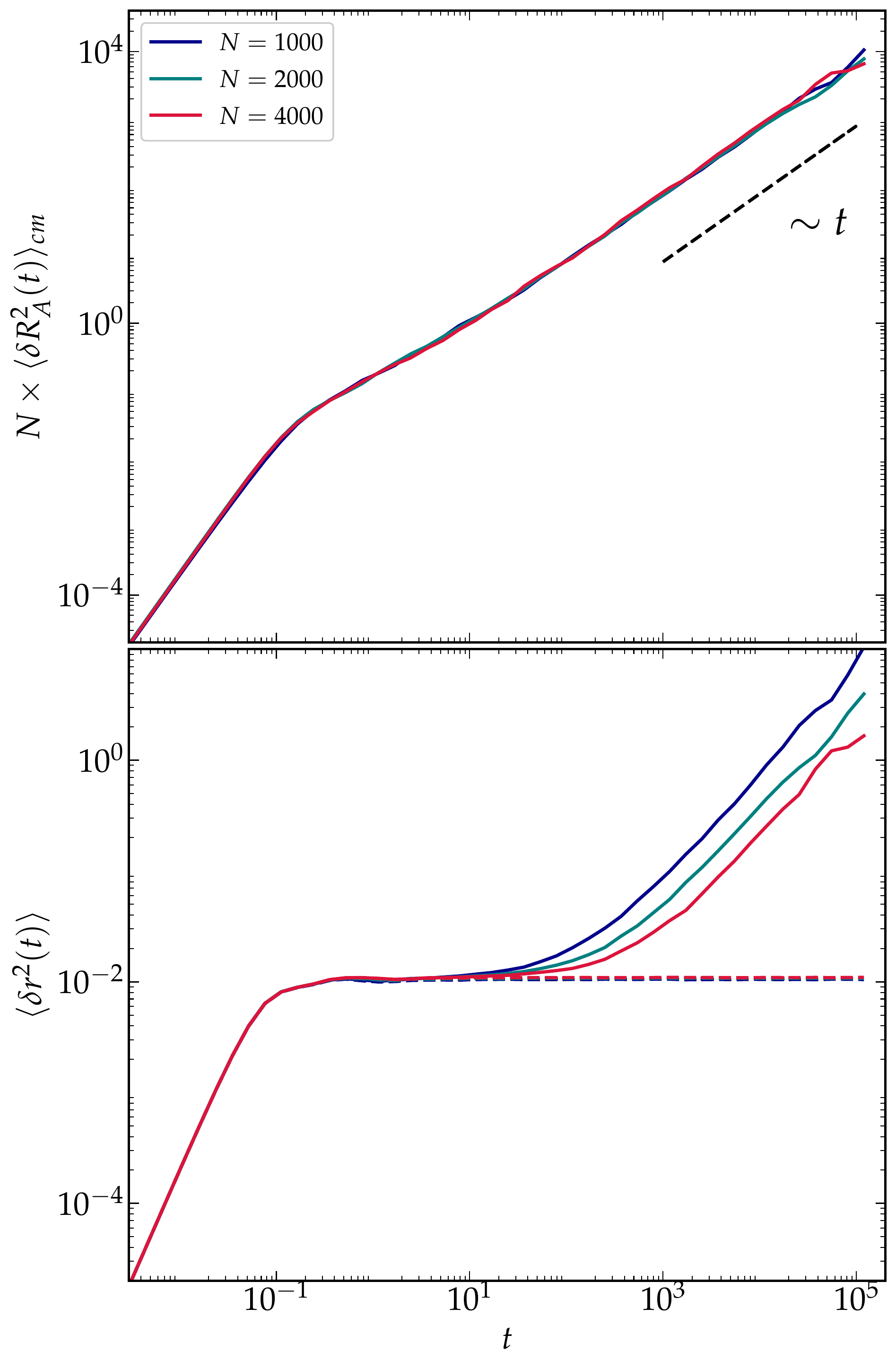}
\caption{{\bf Top panel:} Center-of-mass MSD for A species, scaled with system
size $N$, at density $\rho=2.5$. {\bf Bottom panel:}  Single particle MSD of
the A species at the same density, for the different system sizes.
Solid and dotted lines represent calculations without 
and with center-of-mass correction, respectively.
\label{fig7}}
\end{figure}
Having clarified the actual density dependence of $D_{\rm A}$, we
can now use Eq.~(\ref{eq_dabdarken}) to analyse the behaviour of
the interdiffusion coefficient. If the Manning factor is $S=1.0$,
Eq.~(\ref{eq_dabdarken}) reduces to the Darken equation (\ref{eq_darken}).
Then, when $D_{\rm A}=0$, $D_{\rm AB} \sim D_{\rm B}$, which explains
the observed behaviour of the interdiffusion coefficient in the
regime $\rho \gg \rho_c$.  Physically, even though there is a
dynamical arrest of the A species, because of momentum conservation,
the mobility of B species leads to the center-of-mass motion of the
A species and the consequent finite interdiffusion coefficient. We
also note that even at lower densities, $D_{\rm B} > D_{\rm A}$,
and thus there too, the behaviour of $D_{\rm AB}$ is also dependent
on the diffusive motion of the smaller but faster B particles.

Furthermore, we test the validity of the Darken approximation for
the mixture that we are studying. To do that, we consider two
different quantities, viz $S=D_{\rm AB}/(x_{\rm A} D_{\rm B} +
x_{\rm B} D_{\rm A})$ and $S\Phi$, which are plotted in the bottom
panel of Fig.~\ref{fig5}. If the Darken approximation works, then
$S$ should be around 1. As we can infer from Fig.~\ref{fig5}, over
the whole range of densities, both $S$ and $S \Phi$ deviate from
unity by less than a factor of 2, while the interdiffusion coefficient
decreases by about two orders of magnitude.  So neither the cross
correlations nor the thermodynamic factor strongly affect the density
dependence of $D_{\rm AB}$ which is therefore essentially given by
the linear combination of selfdiffusion coefficients, i.e.~$D_{\rm
AB} \approx 0.5 (D_{\rm A} + D_{\rm B}$).

\begin{figure}[]
\includegraphics[width=7cm]{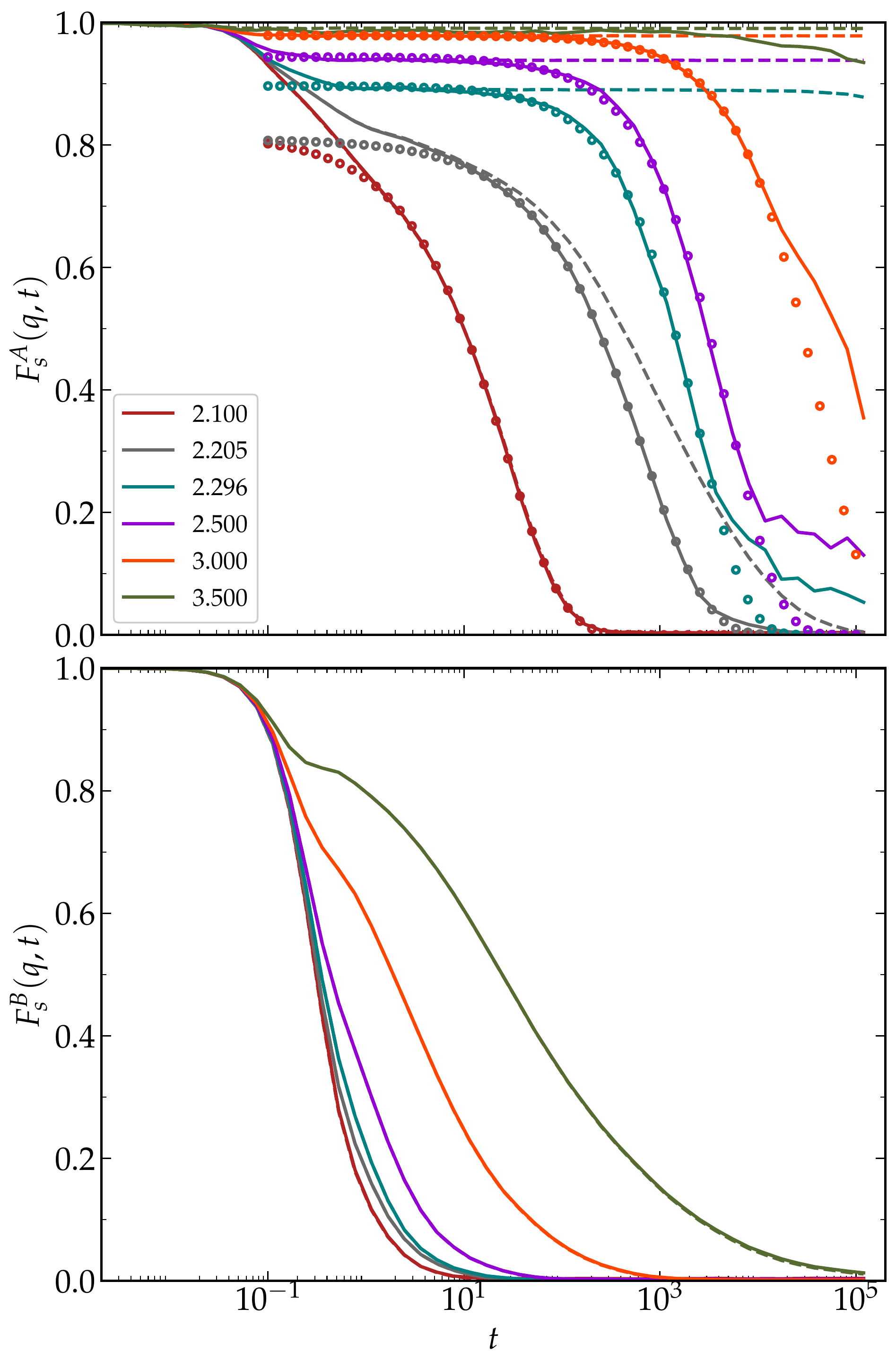}
\caption{Self-intermediate scattering function, $F_s^{\alpha}(q,t)$, for
bigger A (top) and smaller B (bottom) particles, measured at $q=6$, for
different densities. In the top panel, solid and dashed lines
represent calculations without and with
center-of-mass correction, respectively. The dotted lines are fits with
stretched exponentials (see text). The values of the stretching exponent 
are $\beta = 0.73$, 0.68, 0.89, 0.86, and 0.87 for the densities
$\rho = 2.1$, 2.205, 2.296, 2.5, and 3.0, respectively. \label{fig8}}
\end{figure}
In the above discussions, we have noted that the MSD curves for B
species do not show any plateau-like feature, i.e.~the absence of
any signatures of caging by neighbouring particles. Rather, these
curves show anomalous intermediate sub-diffusive dynamics prior to
diffusion. This is similar to what is observed in the case of
interacting particles moving in a quenched environment of soft
obstacles \cite{schnyder2018, schnyder2015}. For the binary mixture
that we are studying, this is a reasonable scenario considering the
large size ratio (see Fig.~\ref{fig1} for the visualisation). At
the large density beyond $\rho_c$ where we are probing the dynamics,
the A population is nearly frozen, and the B particles are diffusing
through this quenched environment with which they interact via some
soft interaction. For the case of interacting particles moving in
a quenched environment of soft obstacles, an avoided localization
transition is observed \cite{schnyder2018}, and one would expect a
similar situation for the eventual dynamical arrest of the B species
at large enough densities \cite{horbach2009}.

Next, we analyse the finite size effects in the measured diffusivities.
Since the interdiffusion coefficient can be written as $D_{\rm AB}
= \Phi L$, where $L = N D_{\rm A}^{\rm cm}$, as discussed in the
previous section, $D_{\rm AB}$ has no finite size effects. This is
illustrated in the top panel of Fig.~\ref{fig7}, where we plot the
center-of-mass MSD of A species scaled with system size, at density
of $\rho=2.5$, for a density larger than $\rho_c$. We observe that
the curves for the different system sizes collapse for the scaled
MSD, implying that ${N}D_{\rm A}^{\rm cm}$ is the same and therefore
$D_{\rm AB}$ remains unaffected. On the other hand, the bare single
particle MSDs show system size dependence at the same density, as
shown in the bottom panel of Fig.~\ref{fig7}. However, if we now
measure the MSD of the particles in the frame of reference of the
center of mass of the A particles, see Eq.~(\ref{eq_rprime}), we
observe that all collapse on the same plateau; see top panel of
Fig.~\ref{fig7}.  This implies, that the finite size dependence in
the single particle dynamics is coming from the change in centre-of-mass
motion. However, as discussed above, this effect gets scaled out
when computing the interdiffusion coefficient.

So far, we have been discussing the anomalous effects in the mean
squared displacement for $\rho > \rho_c$ and the rectification that
is needed. This is also the case in other dynamical observables.
In the top panel of Fig.~\ref{fig8}, we illustrate this for the
incoherent intermediate scattering function $F_s^{\rm A}(q,t)$ at
a wavenumber of $q=6$, corresponding to correlations on nearest-neighbour
distances. The bare measurements are shown with solid lines and the
rectified curves as dashed lines, i.e.~measurements done in the
frame of reference of the center of mass. Consistent with the MSD
data, at low densities (here $\rho = 2.1$), the bare and the rectified
$F_s^{\rm A}(q,t)$ coincide, while at higher densities the rectified
function shows a slower and a more stretched decay (cf.~the curves
for $\rho = 2.205$).  For $\rho \geq \rho_c$, the center-of-mass-corrected
functions show extended plateaus in time, reflecting that in the
frame of reference of the center of mass, the A particles are stuck
in their locations.

The bare curves show a relaxation process, which can be fitted with
stretched exponential functions, $f(q,t)=A_{\rm sef}
\exp[-(t/\tau_q)^\beta]$ (with $A_{\rm sef}$, $\tau_q$, and $\beta$
being fit parameters). The values of the exponent $\beta$, as
obtained from the fits, are in the range between 0.7 and 0.9 (see
caption of Fig.~\ref{fig8}). With increasing density, the stretching
exponent $\beta$ tends to increase towards 1.0, i.e.~towards a
simple exponential decay.  This decay of $F_s^{\rm A}(q,t)$ at
$q=6.0$ reflects the center-of-mass motion of the A species. At
very long times, however, there is the crossover to a much slower
decay, which is due to the fact that there are no rearrangements
in the configuration of A particles. The bottom panel of Fig.~\ref{fig8}
shows $F_s^{\rm B}(q,t)$.  Here, for the low densities, the decay
happens very quickly. Only for $\rho > \rho_c$, the relaxation
timescales start increasing significantly.

\section{Summary and conclusions}
\label{sec4}
Our work elucidates several aspects of the diffusion dynamics in
an equimolar glassforming binary soft-sphere mixture with large
size ratio.  We observe a very pronounced time scale separation
between the motion of the slow A species and that of the fast B
species at high densities. As a consequence, the A particles show
the typical glassy dynamics of a densely packed system, while the
B particles remain mobile at very high densities, $\rho > \rho_c$,
exploring the void space in between the A particles. We have seen
that the interdiffusion coefficient $D_{\rm AB}$ can be well
approximated by the Darken equation (\ref{eq_darken}).  Since at
high density $D_{\rm A} \ll D_{\rm B}$, one therefore obtains $D_{\rm
AB} \approx \Phi x_{\rm A} D_{\rm B}$, i.e.~the interdiffusion
coefficient is dominated by the fast species.  For our equimolar
mixture, we can also write $D_{\rm AB} = \Phi N D_{\rm A}^{\rm cm}$
and thus at high density we have $D_{\rm A}^{\rm cm} \approx D_{\rm
B}/(2N)$.  This implies that even if the A particles form a frozen-in
structure with essentially no rearrangements of the relative positions
of particles, they may follow collectively the diffusive motion of
their center of mass and one finds for the selfdiffusion coefficient
$D_{\rm A} \approx D_{\rm A}^{\rm cm} \propto N^{-1}$.
One can, of course, correct for this finite-size effect by
computing one-particle quantities such as $\langle \delta r^2_{\rm
A}(t) \rangle$ and $F_s^{\rm A}(q,t)$ from the center-of-mass-corrected
coordinates, as defined by Eq.~(\ref{eq_rprime}).

The finite-size effects, that we have reported in this work with respect
to the selfdiffusion coefficient of the slow species, are expected to 
be a typical feature in glassforming binary mixtures with large size
ratio. Moreover, one may expect similar effects in any glassforming 
system with strong dynamic heterogeneities. 
%For example, in a recent simulation study of a metallic glassforming 
%melt \cite{wu2018}, it has been shown that the incoherent intermediate scattering
%function $F_s(q,t)$ of selected particles belonging to a ``rigid''
%icosahedral cluster may exhibit a compressed exponential relaxation,
%as opposed to the stretched exponential decay of $F_s(q,t)$ on 
%average. 

\begin{acknowledgments}
%P. C. thanks his colleague S. Bestiale for useful discussions.
We thank IMSc HPC facility for providing computational resources for our work.
\end{acknowledgments}

%

%%%%%%%%%%%%%%%%%%%%%%%%%%%%%%%%%%%%%%%%%%%%%%%%%%%%%%%%%%%%%%%%%%%%%%%%%%%%

\begin{thebibliography}{99}
%
\bibitem{bechinger2013}
C. Bechinger, F. Sciortino, and P. Ziherl (eds.),
{\it Physics of Complex Colloids} (IOS Press, Amsterdam, 2013).
%
\bibitem{weiss2014}
M. Weiss, in New Models of the Cell Nucleus: Crowding,
Entropic Forces, Phase Separation, and Fractals, edited by 
R. Hancock and K. W. Jeon, International review of cell and molecular 
biology Vol. 307 (Academic Press, San Diego, CA, 2014), Chap. 11, pp. 383–417.
%
\bibitem{hoefling2013}
%Anomalous transport in the crowded world of biological cells
F. H\"ofling and T. Franosch,
Rep. Prog. Phys. {\bf 76}, 046602 (2013).
%
\bibitem{imhof1995_1}
A. Imhof and J. K. G. Dhont, 
Phys. Rev. Lett. {\bf 75}, 1662 (1995).
%
\bibitem{imhof1995_2}
A. Imhof and J. K. G. Dhont, 
Phys. Rev. E {\bf 52}, 6344 (1995).
%
\bibitem{kurita2010}
%Glass transition of two-dimensional binary soft-disk mixtures with large size ratios
R. Kurita and E. R. Weeks,
Phys. Rev. E {\bf 82}, 041402 (2010).
%
\bibitem{blochowicz2012}
%Signature of a Type-A Glass Transition and Intrinsic Confinement Effects in a Binary Glass-Forming System
T. Blochowicz, S. Schramm, S. Lusceac, M. Vogel, B. St\"uhn, P. Gutfreund, and B. Frick,
Phys. Rev. Lett. {\bf 109}, 035702 (2012).
%
\bibitem{bierwirth2018}
%Coexistence of two structural relaxation processes in monohydroxy 
%alcohol–alkyl halogen mixtures: Dielectric and rheological studies
S. P. Bierwirth, C. Gainaru, and R. B\"ohmer,
J. Chem. Phys. {\bf 149}, 044509 (2018).
%
\bibitem{sentjabrskaja2016}
% Anomalous dynamics of intruders in a crowded environment of mobile obstacles
T. Sentjabrskaja, E. Zaccarelli, C. De Michele, F. Sciortino, P. Tartaglia, 
T. Voigtmann, S. U. Egelhaaf, and M. Laurati,
Nat. Commun. {\bf 7}, 11133 (2016).
%
\bibitem{laurati2019}
E. Martinez-Sotelo, M. A. Escobedo-Sánchez, and M. Laurati,
J. Chem. Phys. {\bf 151}, 164504 (2019).
%
\bibitem{moreno2006}
A. J. Moreno and J. Colmenero, 
% Relaxation scenarios in a mixture of large and small spheres: 
% dependence on the size disparity. 
J. Chem. Phys. {\bf 125}, 164507 (2006).
%
\bibitem{horbach2009}
J. Horbach, and T. Voigtmann,
Phys. Rev. Lett. {\bf 109}, 205901 (2009).
%
\bibitem{xu2012}
%Structure, compressibility factor, and dynamics of highly size-asymmetric binary hard-disk liquids
W.-S. Xu, Z.-Y. Sun, and L.-J. An,
J. Chem. Phys. {\bf 137}, 104509 (2012).
%
\bibitem{xu2015}
%Relaxation dynamics in a binary hard-ellipse liquid†	Check for updates
W.-S. Xu, Z.-Y. Sun, and L.-J. An,
Soft Matter {\bf 11}, 627 (2015).
%
\bibitem{lazaro2019}
%Glassy dynamics in asymmetric binary mixtures of hard spheres
E. L\'azaro-L\'azaro, J. A. Perera-Burgos, P. Laermann, T. Sentjabrskaja, 
G. P\'erez-\'Angel, M. Laurati, S. U. Egelhaaf, M. Medina-Noyola, T. Voigtmann, 
R. Castaneda-Priego, and L. F. Elizondo-Aguilera,
Phys. Rev. E {\bf 99}, 042603 (2019).
%
\bibitem{bosse1987}
J. Bosse and J. S. Thakur, 
Phys. Rev. Lett. {\bf 59}, 998 (1987).
%
\bibitem{bosse1995}
J. Bosse and Y. Kaneko, 
%Self-diffusion in supercooled binary liquids. 
Phys. Rev. Lett. {\bf 74}, 4023 (1995).
%
\bibitem{voigtmann2011}
T. Voigtmann, 
%Multiple glasses in asymmetric binary hard spheres. 
Europhys. Lett. {\bf 96}, 36006 (2011).
%
\bibitem{schnyder2018}
S. K. Schnyder and J. Horbach, 
Phys. Rev. Lett. {\bf 120}, 078001 (2018).
%
\bibitem{kurzidim2011}
% Dynamic arrest of colloids in porous environments: disentangling crowding and confinement
J. Kurzidim, D. Coslovich, and G. Kahl,
J. Phys.: Condens. Matter {\bf 23}, 234122 (2011).
%
\bibitem{fitts1962}
D. D. Fitts, 
{\it Non-equilibrium Thermodynamics} 
(MacGraw-Hill, New York, 1962).
%
\bibitem{akcasu1997}
A. Z. Akcasu,
Macromol. Theory Simul. {\bf 6}, 679 (1997).
%
\bibitem{horbach2007} 
J. Horbach, S. K. Das, A. Griesche, M.-P. Macht, G. Frohberg, and A. Meyer,
Phys. Rev. B {\bf 75}, 174304 (2007).
%
\bibitem{darken1949}
L. S. Darken, 
Trans. AIME {\bf 180}, 430 (1949).
%
\bibitem{hartley1949}
G. S. Hartley and J. Crank, 
Trans. Faraday Soc. {\bf 45}, 801 (1949).
%
\bibitem{kuhn2014}
P. Kuhn, J. Horbach, F. Kargl, A. Meyer, and Th. Voigtmann,
Phys. Rev. B {\bf 90}(2), 024309 (2014).
%
\bibitem{bearman1960}
R. J. Bearman, 
J. Chem. Phys. {\bf 32}, 1308 (1960).
%
\bibitem{brochard1986}
F. Brochard and P. G. de Gennes,
Europhys. Lett. {\bf 1}, 221 (1986).
%
\bibitem{sillescu1987}
H. Sillescu,
Makromol. Chem., Rapid Commun. {\bf 8}, 393 (1987).
%
\bibitem{hess1990}
W. Hess, G. N\"agele, and A. Z. Akcasu, 
J. Polym. Sci., Part B: Polym. Phys. {\bf 28}, 2233 (1990).
%
\bibitem{akcasu1991}
A. Z. Akcasu, G. N\"agele, and R. Klein, 
Macromolecules {\bf 24}, 4408 (1991).
%
\bibitem{latz1990}
A. Latz,
{\it Verallgemeinerte konstituierende Gleichungen und Formfaktoren f\"ur
einfache Glasbildner},  
Ph.D. thesis, TU M\"unchen, Germany, 1990.
%
\bibitem{weeks1971}
J. D. Weeks, D. Chandler, and H. C. Andersen, 
J. Chem. Phys. {\bf 54}, 5237 (1971).
%
\bibitem{plimpton1995} 
S. Plimpton,  
J. Comp. Phys. {\bf 117}, 1 (1995).
%
\bibitem{allen2017}
M. P. Allen and D. J. Tildesley,
{\it Computer Simulation of Liquids, 2nd ed.}
(Oxford University Press, Oxford, 2017).
%
\bibitem{soddemann2003} 
T. Soddemann, B. Dünweg, and K. Kremer,
Phys. Rev. E {\bf 68}, 046702 (2003).
%
\bibitem{grigera2001}
T. S. Grigera and G. Parisi,
Phys. Rev. E {\bf 63}, 045102 (2001).
%
\bibitem{berthier2019}
L. Berthier, E. Flenner, C. J. Fullerton, C. Scalliet, and M. Singh,
J. Stat. Mech. 064004 (2019).
%
\bibitem{hansen1986}
J.-P. Hansen and I. R. McDonald,
{\it Theory of Simple Liquids}
(Academic Press, London, 1986).
%
\bibitem{binder2011}
K. Binder and W. Kob,
{\it Glassy Materials and Disordered Solids: An Introduction to Their
Statistical Mechanics, Rev. Ed.}
(World Scientific, Singapore, 2011).
%
\bibitem{manning1961}
J. R. Manning, 
Phys. Rev. {\bf 124}, 470 (1961).
%
\bibitem{steinhardt1983}
P. J. Steinhardt, D. R. Nelson, and M. Ronchetti,
Phys. Rev. B {\bf 28} (2), 784 (1983).
%
\bibitem{dellago2008}
W. Lechner and C. Dellago,
J. Chem. Phys. {\bf 129}, 114707 (2008).
%
\bibitem{schnyder2015}
S. K. Schnyder, M. Spanner, F. H\"ofling, T. Franosch, and J. Horbach, 
Soft Matter {\bf 11}, 701 (2015).
%
%\bibitem{wu2018}
%Z. W. Wu, W. Kob, W.-H. Wang, and L. Xu,
%Nat. Commun. {\bf 9}, 5334 (2018).
%
\end{thebibliography}
\end{document}